\documentclass{ws-spin}
\usepackage{multicol}
\usepackage{graphicx}
\usepackage{latexsym}
\usepackage{amssymb}
\usepackage{amsmath}
\usepackage{amssymb}
\usepackage{tikz}
\usepackage{lipsum}
\begin{document}

\catchline{1}{1}{2022}{}{}
\markboth{Vinod Ashokan; A. Khater and M. Abou Ghantous}{Sublattice magnetizations of nano-magnetic layered materials}

\title{Sublattice magnetizations of ultrathin ferrimagnetic lamellar nanostructures between cobalt leads}

\author{Vinod Ashokan\footnote{Corresponding author}}

\address{Department of Physics,\\ Dr. B. R. Ambedkar National Institute of Technology, \\Jalandhar (Punjab) 144 027, India\\
	\email{ashokanv@nitj.ac.in}}

\author{A. Khater}
\address{ Department of Physics, Le Mans University, 72085 Le Mans, France;\\
	Department of Theoretical Physics, Jan Dlugosz University, Czestochowa, Poland}

\author{M. Abou Ghantous}
\address{Science Department, American University of Technology,\\ Fidar Campus, Halat, Lebanon
}
\maketitle


\begin{abstract}
	In this work we model the salient magnetic properties of the alloy lamellar ferrimagnetic   nanostructures $[Co_{1-c}Gd_c]_{\ell^{\prime}}[Co]_\ell[Co_{1-c}Gd_c]_{\ell^{\prime}}$ between $Co$ semi-infinite leads. We have employed the Ising spin effective field theory (EFT)  to compute the reliable magnetic exchange  constants for the pure cobalt $J_{Co-Co}$ and gadolinium $J_{Gd-Gd}$ materials, in complete agreement with their experimental data. The sublattice magnetizations of the $Co$ and $Gd$ sites on the individual hcp atomic (0001) planes of the $Co-Gd$ layered nanostructures are computed  for each plane and corresponding sites, by using the combined EFT and mean field theory (MFT) spin methods. The sublattice magnetizations,  effective site magnetic moments, and ferrimagnetic compensation characteristics for the individual hcp atomic planes of the embedded nanostructures, are computed  as a function of temperature, and for various stable eutectic concentrations in the range $c\leq$ 0.5. The theoretical results for the sublattice magnetizations and the local magnetic variables of these ultrathin ferrimagnetic lamellar nanostructured systems, between cobalt leads, are necessary for the study of their magnonic transport properties, and eventually their spintronic dynamic computations. The method developed in this work is general and can be applied to comparable magnetic systems nanostructured with other materials.
\end{abstract}


\begin{multicols}{2}
\section{Introduction}

The multilayered lamellar nano-magnetic  nanostructure has made a tremendous progress in preparing and analyzing the physical properties of nano-magnetic layered nanostructures and magnetic  nano-junctions. These systems have a panoply of industrial and technological applications in the areas of spin wave magnonics \cite{Serga07,Schneider08,Kruglyak10,Lee08}, and spintronics \cite{Wolf04,Zutic04,Bogani08}. However, the study of nano-magnetic lamellar multilayered nanostructure and nanojunction composite of rare earth-transition metal alloy systems are still in its infancy. A fundamental and intriguing interest associated with such rare earth-transition metal systems is to understand the phenomena which may arise due to the decrease of their size when surface and quantum mechanical effects come into play. The atomic interfaces and sublattice magnetization in these nanostructures are become critical components for the physics of embedded nano-junctions.


The nano-magnetic properties of ferrimagnetic alloy nanostructure of  rare earth-transition metal has been reported in the literature \cite{Camley93,Andres00,Anilturk03,Gonzalez04,Andres08,Demirtas05,Javier22}. The nano fabrication technique has made the experimental progress to realize the thin multilayered nanostructures with novel physical properties and promising applications in magnonic devices. In this  work we study in particular the Co-Gd rare earth-transition metal alloy system; the transition metal  $Co$ and rare earth $Gd$ are ferromagnetic with their Curie temperatures of 119.4 and 25.2 meV respectively. Nano-magnetic nanojunctions build from Co-Gd alloys in diverse multilayer formats can present hence  very useful properties for technological applications at room temperature. In this respect, the  $bulk$ $Co_{1-c}Gd_{c}$ alloy materials with different alloy concentrations $c$, have been studied intensively in the past for diverse applications in  sensors, magneto optical devices and magnetic storage elements \cite{Chaudhari73,Hansen89}.

A greater understanding of lamellar  Co/Gd multilayers nanostructures has been achieved \cite{Andres00,Gonzalez04,Andres08}. It is noted that when we form an amorphous alloy $Co_{1-c}Gd_{c}$ in these systems there is a strong asymmetric spontaneous diffusion of $Co$ into the $Gd$ plane occurs, and interfaces for various concentration. The experimental techniques made possible to control the interdiffusion  for some stable eutectic concentration $c\leq 0.5$ while  preserving the ferrimagnetic structure of the multilayer systems. The properties of multilayers presenting alloy interfaces, with few lamellar atomic planes thick significantly depends on the   degree of material interdiffusion \cite{Andres00,Gonzalez04}. Such interdiffusion may play crucial role in determining  the nano-magnetic properties of the multilayers systems \cite{Camley93}, because the individual planes are having different exchange couplings at their interfaces from the bulk. This has also been reported earlier by model calculations which show that nano-atomic scale magnetic alloyed interfaces can significantly modify the nano-magnetic properties of multilayer systems \cite{Khater92,Fresneau94,Khater021}. The preparation of alloy like and composition stable nanojunctions composed of $Co$ and $Gd$ between cobalt leads is hence possible in principle experimentally due to the method of controlled interdiffusion process.

It should be noted that there has been attempts in the past to model the magnetic properties of bulk and layered cobalt-gadolinium systems \cite{Demirtas05,Camley88,Mansuripur86} using the MFT method. These model calculations have been performed by adjusting in general the MFT results to fit the experimental data, using the cobalt spin $S_{Co}$  as a fitting parameter. In some of the calculations where $S_{Co}$ is assigned its fundamental value, the overall fit function for the magnetization with temperature does not give better agreement with the experimental data. Furthermore, the asymmetric choice of nearest neighbor exchange constants for cobalt $J_{Co-Co}$ and gadolinium $J_{Gd-Gd}$ is made in these references  to reduce the number of adjustable parameters, but without giving any fundamental justification. To add to this complex situation, there is a wide array of experimental values of exchange constant for the cobalt $J_{Co-Co}$ and gadolinium $J_{Gd-Gd}$ are available in the literature from different types of measurements \cite{Vaz08}, which does not help to clarify the situation for advanced modeling.

In our previous work \cite{Ashokan14}, we computed the  ballistic and scattering  transport properties of spin waves (SW) incident from cobalt leads, on to the embedded ultrathin ferrimagnetic cobalt-gadolinium $[Co_{1-c}Gd_{c}]_\ell$ nanojunction systems between the leads. The nanojunction $[Co_{1-c}Gd_{c}]_\ell$ is principle a randomly disordered alloy with varied hcp atomic palnes $\ell$ between matching hcp planes of the $Co$ leads, at known stable concentrations  $c\leq 0.5$ for this nano-alloy system. To be able to carry out these computations it was necessary to compute the sublattice magnetizations and magnetic exchange constants in this system \cite{Ghantous13}.

In the present work we have modeled the sublattice magnetizations and magnetic exchange constants of the alloy layered nanostructures $[Co_{1-c}Gd_c]_{\ell^{\prime}}[Co]_\ell[Co_{1-c}Gd_c]_{\ell^{\prime}}$ sandwiched between semi-infinite cobalt leads, at concentrations  $c\leq 0.5$. These triple-nanostructure systems are more complex than the previously studied single-nanostructure systems \cite{Ghantous13}. The present work is hence motivated by the objective to present a more complex computational modeling of the salient nano-magnetic properties of the triple-nanostructures for fundamental interest, and the  ballistic transport and scattering of spin waves incident from cobalt leads on complex ferrimagnetic cobalt-gadolinium nanojunction systems \cite{Ashokan15}. Such complex systems have as it turns out a richer and wider range of spin wave filtering properties. The complex embedded triple-nanostructures are denoted symbolically henceforth by $[\ell^{\prime}\ell \ell^{\prime}]$ for convenience, corresponding to the alternating alloy and pure nanostructures. The basal hcp (0001) atomic planes of the $...Co][\ell^{\prime}\ell \ell^{\prime}][Co...$  nanojunction systems are normal to the direction of the c-axis itself considered to be along the direction of the leads.

The alloy layered $[\ell^{\prime}\ell \ell^{\prime}]$ nanostructures under consideration are ultrathin $\sim 1.5$ nm composite cobalt-gadolinium alloy systems sandwiched between Co leads, and are hence different from bulk alloy and multilayer systems \cite{Camley93,Mansuripur86,Demirtas05}. Also, due to the absence of  first principle calculations for Co-Co and Co-Gd exchange in the alloy layered $[\ell^{\prime}\ell \ell^{\prime}]$ nanostructures.There is hence effectively a need for reliable data for the exchange and sublattice magnetizations in these systems to be able to develop  modeling studies for the $...Co][\ell^{\prime}\ell \ell^{\prime}][Co...$  nanojunctions which are key elements for ballistic spin wave transport in magnonic devices \cite{Ashokan14,Ashokan16,Khater18,Chelli20}. This need has motivated our EFT calculations to determine such data with no fitting parameters, using basic values $S_{Co}$ = 1 and $S_{Gd}$ = 7/2 as the spin references at absolute 0 $K$.

The structure of the paper is as follows. In section \ref{EFTmodel},  the EFT Ising spin method with experimental data is computed for the  reliable $J_{Co-Co}$ and $J_{Gd-Gd}$ exchange for the pure crystalline cobalt and gadolinium materials. These are then attributed to nearest neighbor $Co-Co$ and $Gd-Gd$ interactions in the $...Co][\ell^{\prime}\ell \ell^{\prime}][Co...$   nanojunction systems for eutectic stable concentrations $c\leq 0.5$. The  combined EFT and MFT methods are presented in section \ref{seededmftcal}, to compute sublattice magnetizations for the cobalt and gadolinium sites on the individual hcp basal atomic planes of the alloy layered $[\ell^{\prime}\ell \ell^{\prime}]$ lamellar nanostructures as a function of temperature, with thicknesses $[2'22']$ and $[3'33']$, and for different alloy concentrations $c$. The sublattice magnetizations and corresponding ferrimagnetic compensation temperatures are shown in this section. The overall discussions and conclusions are presented in section \ref{Conclusions}.

\begin{figure*}
	\centering
	\includegraphics[scale=0.9]{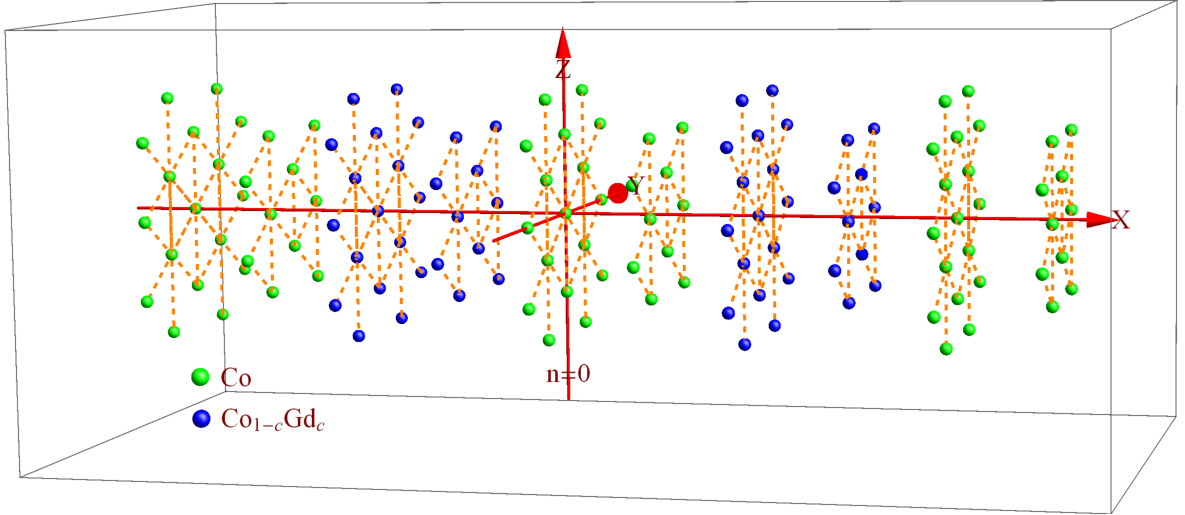}
	\caption{Schematic representation for the consecutive planes of cobalt-gadolinium VCA alloy and pure cobalt plane of nanojunction  $[Co_{1-c}Gd_c]_{2}[Co]_2[Co_{1-c}Gd_c]_{2}$ between crystalline cobalt leads. The hcp crystal
		c-axis is normal to the symmetry (0001) atomic planes.}\label{fig}
\end{figure*}

\section{EFT modeling of pure $Co$ and $Gd$ systems}
\label{EFTmodel}

The  EFT modeling  incorporates the contribution of the single site spin correlations to the order parameter, and hence known to be superior to the MFT.  We use it in the present work to model and calculate the exchange constant for $Co$ and $Gd$ crystals over their ordered nano-magnetic phase, by comparing the EFT magnetization results and Curie temperatures with the experimental data \cite{Kuzmin05}. The exchange for $Co$ and $Gd$ crystals are then calculated by using our EFT constitutive relations \cite{Khater15} $kT_c/zJS(S+1) = 0.3127$ and $kT_c/zJS(S+1) = 0.3162$, valid for cobalt and gadolinium, respectively. The EFT exchange $J_{Co-Co}$ and $J_{Gd-Gd}$ are in good agreement with the mean values of exchange constant obtained from extensive experimental data \cite{Vaz08} for $Co$ and $Gd$. The calculated EFT magnetization and exchange results for cobalt are then used for the calculations of sublattice magnetizations of the alloy layered $...Co][\ell^{\prime}\ell \ell^{\prime}][Co...$  nanostructure between Co leads, by using the MFT method \cite{Camley88,Mansuripur86,Demirtas05}. The schematic representation for the consecutive planes of cobalt-gadolinium VCA alloy and pure cobalt plane of nanojunction  $[Co_{1-c}Gd_c]_{2}[Co]_2[Co_{1-c}Gd_c]_{2}$ between crystalline cobalt leads are shown in Fig.\ref{fig}.

The Ising spin Hamiltonian $\mathcal{H}$ in the absence of local spin anisotropy and Zeeman effects may be expressed as,
\begin{equation}
	\label{IsingspinH}
	\mathcal{H}=-J\sum_{<i,j>}S_{iz}\,.\,S_{jz},
\end{equation}
where, $\langle i, j\rangle$ represents sum over nearest neighbors in the crystal, and $J$ is the nearest neighbors magnetic exchange constant that induces spin order along a selected z-axis. The coordination number of  $Co$ and $Gd$ is  $z$ = 12, and present negligible anisotropy in the bulk material as compared to the exchange. The Hamiltonian Eq.(\ref{IsingspinH}) for computational purposes may be written in more useful form as,

\begin{equation}
	\mathcal{H}=\sum_{i}{\sum_{j}(-J S_{jz})\;S_{iz}}\equiv \sum_i{\mathcal{H}_{i}(x)}\equiv \sum_i{-x \;S_{iz}},
\end{equation}

The thermodynamic canonical averages for any desired \textit{spin operator} ``$\mathrm{Op}$'' may be calculated by the effective filed theory method using the Van der Wearden's (VdW) operator $\exp\left(J S_z\nabla\right)$. The Mathematica code formulation for any given characteristic function $f_{\mathrm{Op}}(x)$ of the spin system may be expressed as

\begin{eqnarray}
	\label{genfunction}
	f_{\mathrm{Op}}(x)=\frac{\mathrm{Tr} (\mathrm{Op.MatrixExp}[- \mathcal{H}_{i}(x)/kT])}{\mathrm{Tr} (\mathrm{MatrixExp}[-\mathcal{H}_{i}(x)/kT])},
\end{eqnarray}
where the Van der Wearden's (VdW) operator for cobalt with $S=1$ is given as \cite{Tucker94},
\begin{equation}
	\exp\left(J S_z\nabla\right)=S_z^{2}\cosh\left(J\nabla\right)+S_z\sinh\left(J\nabla\right)+1-S_z^{2}.
\end{equation}
The differential operator $\nabla=\partial/\partial{x}$ operate with property $ <\exp\left( a\nabla\right)>f_{\mathrm{Op}}(x)|_{x\rightarrow 0}=f_{\mathrm{Op}}(x+a)|_{x\rightarrow 0}=f_{\mathrm{op}}(a)$.
The EFT method calculations are discussed in detail in the earlier works   \cite{Tucker94,Khater11,MAbouGhantous11,Honmura79}. With the help of Matrix quantum mechanics all required averages can be evaluated with symbolic and numerical procedures.

The thermodynamic canonical averages are represented by $<\mathrm{Op}>\,=\,<(\exp\left(J S_z\nabla\right))^z>f_{\mathrm{Op}}(x)|_{x\rightarrow 0}$. The canonical averages are confined to single-site spin variables  $\sigma=<S_z>$ and $q=<S^2_z>$  with reference to their basic spin values $S=1$ for $Co$ and $S=7/2$ for $Gd$ at $T=0\,K$. In the computational procedure desired decoupling approximation is used, which is equivalent to neglecting the  site-site correlations $<S_{iz}S_{jz}>$ but preserving the single-site correlations $<S_{iz}S_{iz}>$. Furthermore, this decoupling approximation makes EFT approach quite effective to compute the salient magnetic properties of the systems.

The comparison between the normalized magnetizations for the $Co$ and $Gd$ crystals calculated using EFT method, and by using the EFT constitutive relations \cite{Khater15}, $kT_c/zJS(S+1) = 0.3127$ for $Co$  ($S=1$, $z=12$), and  $kT_c/zJS(S+1) = 0.3162$ for $Gd$ ($S=7/2$, $z=12$), with their experimental measurements \cite{Kuzmin05,Myers51,Pauthenet82,Nigh63}, yield their respective Curie temperatures, that is 119.4 meV and 25.2 meV  \cite{Ghantous13}. It is striking to note that for $Co$ with spin $S=1$, the EFT calculated magnetization and Curie temperature agree with the experimentally observed mean value of exchange constant  $J_{Co-Co}$ = 15.9 meV  \cite{Vaz08}. Similarly, the hcp $Gd$ with the spin $S=7/2$, the EFT calculated magnetization and Curie temperature agree with the corresponding experimentally measured  exchange $J_{Gd-Gd}$ = 0.42 meV. \cite{Kuzmin05}. The EFT calculated exchange  $J_{Co-Co}$ and spin variable $\sigma_{Co}$ for sites on the cobalt leads seed from the interfaces inwards, to calculate the sublattice magnetization using MFT of embedded alloy layered $[\ell^{\prime}\ell \ell^{\prime}]$ nanostructure.

\section{Sublattice magnetizations of the alloy layered $[\ell^{\prime}\ell \ell^{\prime}]$  ferrimagnetic nanostructures between cobalt leads}
\label{seededmftcal}

In the present computational modeling the alloy hcp atomic planes of the layered  $[\ell^{\prime}\ell \ell^{\prime}]$ nanostructure sandwiched between cobalt leads, are modeled as crystalline atomic planes. On their hcp lattice there is a random homogeneous distributions of $Co$ and $Gd$ atoms. Any random site is considered to have the usual six nearest neighbors in its hcp (0001) basal plane, and another six neighbors on the two adjacent planes. The system is made of alternating hcp (0001) atomic planes, and the structural morphology of the two interfaces between the leads and the layered nanostructure are abrupt and crystalline. The advent of advanced experimental techniques for $Co/Gd$ multilayer systems, \cite{Andres08,Gonzalez04}, permits minimizing the interface roughness, and the control of the atomic interdiffusion towards stable eutectic compositions $c$ = 0.1 to 0.5.

To calculate the sublattice magnetization for the individual basal atomic hcp planes of the alloy layered $[\ell^{\prime}\ell \ell^{\prime}]$ nanostructures by MFT, the Brillouin's functions are used to calculate initially the different spin variables $\sigma^{(n^{\prime})}_\alpha$ for the $n^{\prime}$th layered atomic plane. To simplify the notation we systematically call  $\sigma^{(n^{\prime})}_\alpha$ as the thermodynamic canonical spin variable so that  $\sigma^{(n^{\prime})}_\alpha=S_\alpha.B_S\equiv B_\alpha(S_\alpha,T,H^{(n^{\prime})}_\alpha)$, where $S_\alpha$ represents the fundamental atomic spin and $\alpha$ namely for the Co and Gd atoms, and $B_S$ is the Brillouin function. The thermodynamic canonical spin variable is given as, 
\begin{align}
\label{BFSpinVar}
&\sigma^{(n^{\prime})}_\alpha= \nonumber\\
&\frac{2S_\alpha+1}{2}\mathrm{Coth}\left(\frac{2S_\alpha+1}{2S_\alpha} \frac{H^{(n^{\prime})}_\alpha}{kT}\right)- \frac{1}{2}\mathrm{Coth}\left(\frac{1}{2S_\alpha}\frac{H^{(n^{\prime})}_\alpha}{kT}\right)\nonumber \\
\end{align}
where $H^{(n^{\prime})}_\alpha$ represents the molecular field energy  for the element $\alpha$ in the atomic plane $n^{\prime}$ due to its interaction with its $z$ = 12 nearest neighbors. The $kT$ represents  thermal energy and the effective magnetic moment  per site is $\bar{M}^{(n^{\prime})}$  in the $n^{\prime}$th plane, is given as in units of Bohr magnetons by
\begin{equation}
\label{effecmagmoment}
\bar{M}^{(n^{\prime})}/ \mu_B =(1-c) g^{(n^{\prime})}_{Co} \sigma^{(n^{\prime})}_{Co}+ c g^{(n^{\prime})}_{Gd} \sigma^{(n^{\prime})}_{Gd}.
\end{equation}
$g^{(n^{\prime})}_{\alpha}$ are the g factors for the alloy element on the $n^{\prime}$th plane. The magnetization for the $n^{\prime}$th atomic plane is calculated by multiplying $\bar{M}^{(n^{\prime})}$ by the number of sites per unit volume for the atomic plane.

\subsection{Alloy layered $[Co_{1-c}Gd_{c}]_2[Co]_2[Co_{1-c}Gd_{c}]_2$ ferrimagnetic nanostructure between cobalt leads}
\label{222}
Consider in this subsection the embedded layered $[2^{\prime}2 2^{\prime}]$ nanostructure between cobalt leads. For the $Co$ and $Gd$ atom in the ferrimagnetic alloyed atomic planes are found with the respective probabilities $(1-c)$ and $c$. The molecular field energy using MFT for a $Co$ atom on the 1st hcp basal plane of the layered nanostructure at the interface with the cobalt lead, may hence be expressed as

\begin{eqnarray}
\label{HfieldCo1}
H^{(1)}_{Co} &=& (3 \sigma^{(B)}_{Co} J_{cc})+ 6[(1-c)\sigma^{(1)}_{Co}J_{cc}+c\sigma^{(1)}_{Gd}J_{cg}]\nonumber\\& &+ 3[(1-c)\sigma^{(2)}_{Co}J_{cc}+c\sigma^{(2)}_{Gd}J_{cg}].
\end{eqnarray}

The exchange interactions are denoted by the simplified notation $J_{CoCo}\equiv J_{cc}$, $J_{GdGd}\equiv J_{gg}$, and $J_{CoGd}\equiv J_{cg}$. Equally, the molecular field energy for a $Gd$ atom on the 1st hcp basal plane of the alloy layered nanostructure at the interface with the cobalt lead, is

\begin{figurehere}
	\centering
	\includegraphics[scale=0.7]{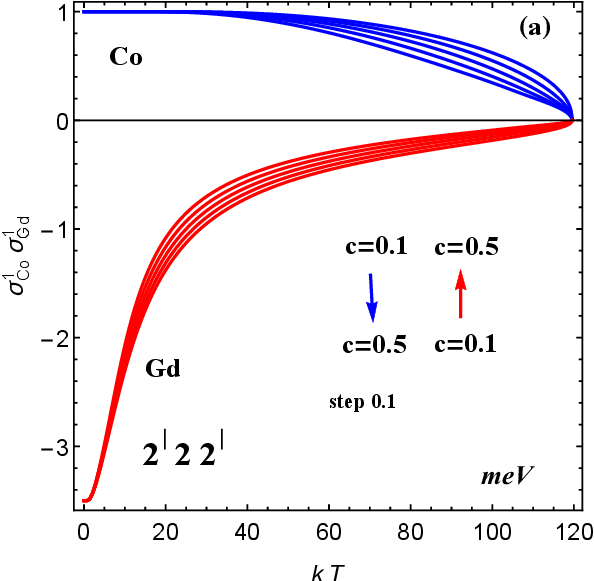}
	\includegraphics[scale=0.75]{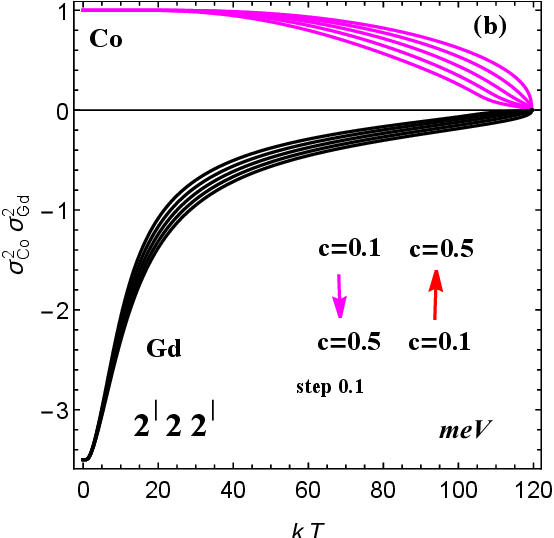}
	\caption{Calculated spin variables $\sigma_{Co}$ and $\sigma_{Gd}$, for $Co$ and $Gd$ sites on the 1st (a) and 2nd (b) hcp basal (0001) planes, of the alloy layered $[Co_{1-c}Gd_{c}]_2[Co]_2[Co_{1-c}Gd_{c}]_2$ ferrimagnetic nanostructure between cobalt leads, for different alloy concentrations $c$, as a function of \textit{kT} in meV. The down (up) arrows in each figure correspond to the trend of the  $\sigma$ spin variations for the $Co$\, ($Gd$) sites with $c$ step changes.}\label{222SpinVarCoGd12}
\end{figurehere}
\begin{eqnarray}
\label{HfieldGd1}
H^{(1)}_{Gd} &=& (3 \sigma^{(B)}_{Co} J_{cg})+ 6[(1-c)\sigma^{(1)}_{Co}J_{cg}+c\sigma^{(1)}_{Gd}J_{gg}]\nonumber\\& &+ 3[(1-c)\sigma^{(2)}_{Co}J_{cg}+c\sigma^{(2)}_{Gd}J_{gg}].
\end{eqnarray}

In the present formulation, the seeding spin value for the lead $Co$ atom at the interface with the alloy layered nanostructure is represented by $\sigma^{(B)}_{Co}$, which is obtained singularly from the EFT calculations described in detail in section \ref{EFTmodel}.

In contrast, the molecular field for a $Co$ atom on the 2nd hcp basal plane of the layered nanostructure inwards from the 1st, is
\begin{eqnarray}
\label{HfieldCo2}
H^{(2)}_{Co} &=& 3[(1-c)\sigma^{(1)}_{Co}J_{cc}+c\sigma^{(1)}_{Gd}J_{cg}]+ 6[(1-c)\sigma^{(2)}_{Co}J_{cc}\nonumber\\& &+c\sigma^{(2)}_{Gd}J_{cg}]+3 \sigma^{(3)}_{Co} J_{cc}
\end{eqnarray}
Similarly, the corresponding molecular field for a $Gd$ atom on the 2nd hcp basal plane of the layered nanostructure, is
\begin{eqnarray}
\label{HfieldGd1}
H^{(2)}_{Gd} &=& 3[(1-c)\sigma^{(1)}_{Co}J_{cg}+c\sigma^{(1)}_{Gd}J_{gg}]+ 6[(1-c)\sigma^{(2)}_{Co}J_{cg}\nonumber\\& &+c\sigma^{(2)}_{Gd}J_{gg}]+3 \sigma^{(3)}_{Co} J_{cg}
\end{eqnarray}

\begin{figurehere}
\centering
\includegraphics[scale=0.60]{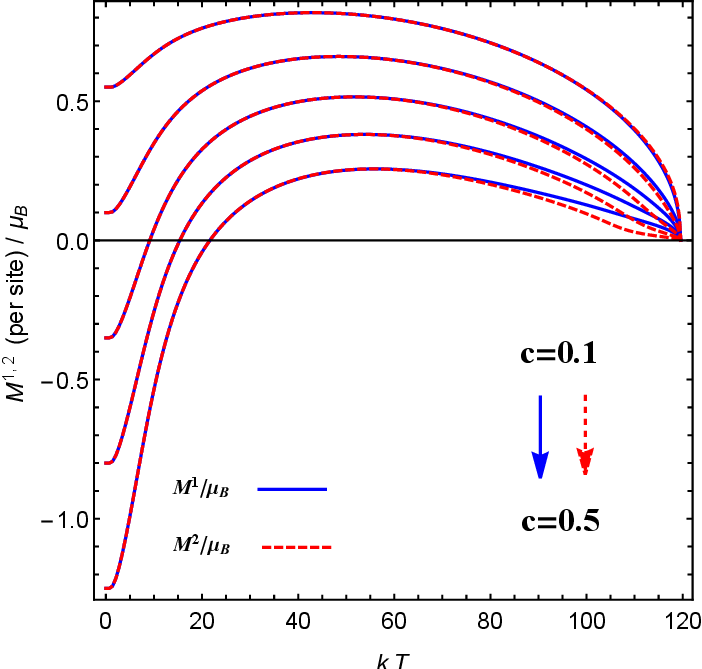}
\caption{Calculated magnetic moments per site for sites on   the 1st (solid curves) and 2nd (dotted curves) hcp basal (0001) planes of the alloy layered ferrimagnetic nanostructure $[Co_{1-c}Gd_{c}]_2[Co]_2[Co_{1-c}Gd_{c}]_2$ between cobalt leads. They present small differences only at the high temperature $kT$ end of the ordered phase. The down arrows follow the variation trend for the magnetic moments per site with the $c$ step changes, on the 1st (solid arrow) and 2nd (dotted  arrow) hcp basal (0001) planes.}\label{222Mmomentpersite12}
\end{figurehere}

The hcp basal planes of the pure $[Co]_2$ layer between the alloy layers $[Co_{1-c}Gd_{c}]_2$, are designated respectively as the 3rd and 4th atomic planes. Using the symmetry properties of the layered $[2^{\prime}2 2^{\prime}]$ nanostructure, we note that $\sigma^{(3)}_{Co}\equiv\sigma^{(4)}_{Co}$. The corresponding molecular field for a $Co$ atom on the 3rd hcp basal plane is hence
\begin{eqnarray}
\label{HfieldCo3}
H^{(3)}_{Co} &=& 3[(1-c)\sigma^{(2)}_{Co}J_{cc}+c\sigma^{(2)}_{Gd}J_{cg}]+9 \sigma^{(3)}_{Co} J_{cc}
\end{eqnarray}

The above equations can be put into matrix form
\begin{align}
\label{nonlineareq222}
&\left(
\begin{array}{c}
	H_{Co}^{(1)}  \\
	H_{Gd}^{(1)}  \\
	H_{Co}^{(2)}  \\
	H_{Gd}^{(2)}  \\
	H_{Co}^{(3)}  \\
\end{array}\right)=
\left(
\begin{array}{c}
	A_1  \\
	A_2\\
	0  \\
	0\\
	0\\
\end{array}\right)+
\left(
\begin{array}{ccccc}
	x_1 & x_2  & x_3 & x_4 & x_5 \\
	y_1 & y_2  & y_3 & y_4 & y_5 \\
	u_1 & u_2  & u_3 & u_4 & u_5 \\
	v_1 & v_2  & v_3 & v_4 & v_5 \\
	z_1 & z_2  & z_3 & z_4 & z_5 \\
\end{array}\right)
\left(
\begin{array}{c}
	\sigma^{(1)}_{Co}  \\
	\sigma^{(1)}_{Gd}  \\
	\sigma^{(2)}_{Co}  \\
	\sigma^{(2)}_{Gd}  \\
	\sigma^{(3)}_{Co}  \\
\end{array}\right)
\end{align}
and the coefficients matrix is identical to
\begin{align}
&\left(
\begin{array}{ccccc}
	x_1 & x_2  & x_3 & x_4 & x_5 \\
	y_1 & y_2  & y_3 & y_4 & y_5 \\
	u_1 & u_2  & u_3 & u_4 & u_5 \\
	v_1 & v_2  & v_3 & v_4 & v_5 \\
	z_1 & z_2  & z_3 & z_4 & z_5 \\
\end{array}\right)&\nonumber\\
&\equiv\left(
\begin{array}{ccccc}
	2 \beta J_{cc} & 6cJ_{cg} & \beta J_{cc}  & 3cJ_{cg} & 0  \\
	2 \beta J_{cg} & 6cJ_{gg} & \beta J_{cg}  & 3cJ_{gg} & 0  \\
	\beta J_{cc} & 3cJ_{cg} & 2\beta J_{cc}  & 6cJ_{cg} & 3J_{cc}   \\
	\beta J_{cg} & 3cJ_{gg} & 2\beta J_{cg}  & 6cJ_{gg} & 3J_{cg}   \\
	0 & 0 & \beta J_{cc}  & 3cJ_{cg} & 9J_{cc}   \\
\end{array}\right)
\end{align}
where $A_1=3 \sigma^{(B)}_{Co} J_{cc}$,  $A_2=3 \sigma^{(B)}_{Co} J_{cg}$ and $\beta=3(1-c)$. Using   Eqs.(\ref{HfieldCo1}) to (\ref{HfieldCo3}), and the spin variables $\sigma^{(n^{\prime})}_{\alpha}$ format given by Eq.(\ref{BFSpinVar}), it follows that Eq.(\ref{nonlineareq222}) represents a  nonlinear equations and to be solved for the spin variables

\begin{figurehere}
	\centering
	\includegraphics[scale=0.58]{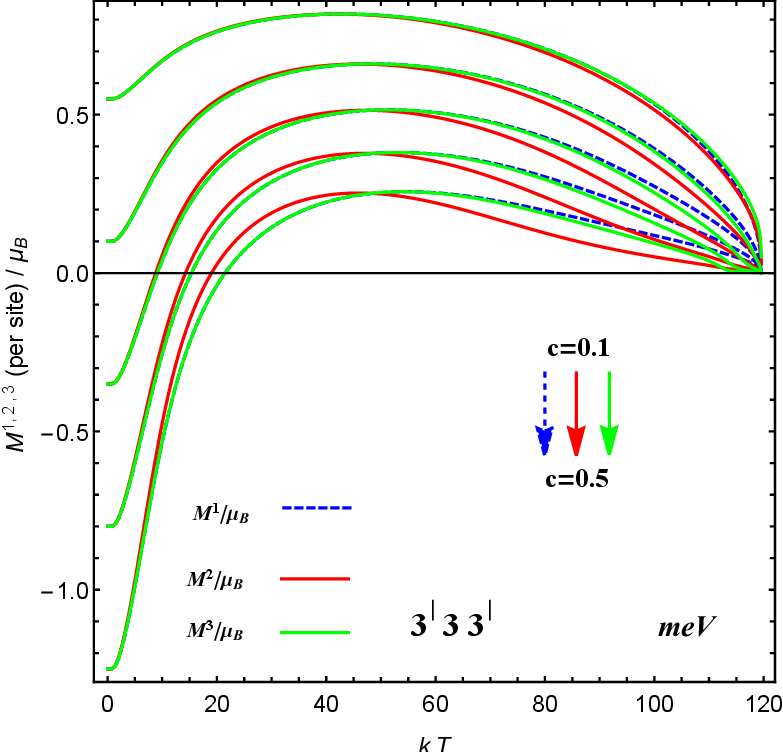}
	\caption{Calculated magnetic moments per site for the 1st, 2nd, and 3rd hcp basal atomic planes of the alloy layered $[Co_{1-c}Gd_{c}]_3[Co]_3[Co_{1-c}Gd_{c}]_3$ ferrimagnetic nanostructure. The magnetic moments per site on the 1st and 3rd hcp planes (discontinuous and continues curves, respectively) are quite similar throughout the temperature range of the ordered phase. They differ significantly from the magnetic moments on the 2nd hcp plane (continues curves). The down arrows correspond to the trend of the magnetic moment variations with $c$ step changes.}\label{333Mmomentpersite123}
\end{figurehere}
\begin{eqnarray}
\left(
\begin{array}{c}
	\sigma^{(1)}_{Co}  \\
	\sigma^{(1)}_{Gd}  \\
	\sigma^{(2)}_{Co}  \\
	\sigma^{(2)}_{Gd}  \\
	\sigma^{(3)}_{Co}  \\
\end{array}\right)=
\left(
\begin{array}{c}
	B_{Co}(S_{Co},T,H_{Co}^{(1)} )  \\
	B_{Gd}(S_{Gd},T,H_{Gd}^{(1)})  \\
	B_{Co}(S_{Co},T,H_{Co}^{(2)})  \\
	B_{Gd}(S_{Gd},T,H_{Gd}^{(2)})  \\
	B_{Co}(S_{Co},T,H_{Co}^{(3)})  \\
\end{array}\right)
\end{eqnarray}

Solving the above equations numerically, we calculate the spin variables $\sigma^{(1)}_{Co}$, $\sigma^{(1)}_{Gd}$, $\sigma^{(2)}_{Co}$, $\sigma^{(2)}_{Gd}$ and $\sigma^{(3)}_{Co}$ as a function of temperature, for any given alloy concentration $c$. Note that the symmetry of the system imposes the following equalities for the spin variables

\begin{eqnarray}
&\left(
\begin{array}{cc}
	\sigma^{(1)}_{Co} \\
	\sigma^{(1)}_{Gd} \\
\end{array}\right)\equiv
\left(
\begin{array}{cc}
	\sigma^{(6)}_{Co}  \\
	\sigma^{(6)}_{Gd}  \\
\end{array}\right),\left(
\begin{array}{cc}
	\sigma^{(2)}_{Co} \\
	\sigma^{(2)}_{Gd} \\
\end{array}\right)\equiv
\left(
\begin{array}{cc}
	\sigma^{(5)}_{Co}  \\
	\sigma^{(5)}_{Gd}  \\
\end{array}\right),\nonumber\\
&\left(
\begin{array}{cc}
	\sigma^{(3)}_{Co} \\
\end{array}\right)\equiv
\left(
\begin{array}{cc}
	\sigma^{(4)}_{Co}  \\
\end{array}\right)
\end{eqnarray}

The calculated spin variables $\sigma^{(n^{\prime})}_{Co}$, $\sigma^{(n^{\prime})}_{Gd}$ for the nominal $n^{\prime}=1$ and $n^{\prime}=2$ hcp basal planes are presented in Fig.\ref{222SpinVarCoGd12}, as a function of temperature and for eutectic concentrations $c=[0.1,0.5]$ in steps of 0.1.

Further, Eq.(\ref{effecmagmoment}), and $g^{(n^{\prime})}_{Co} \equiv g_{Co}$ = 2.2, $g^{(n^{\prime})}_{Gd} \equiv g_{Gd}$ = 2 for all $n{^\prime}$, yield the magnetic moments per site on the 1st and 2nd hcp basal planes as a function of temperature. These are presented for comparison in Fig.\ref{222Mmomentpersite12}, where a small interesting difference is observed  at the high temperature $kT$ end of the ordered phase. Compensation temperatures $kT_{comp}\,<\,21$ meV, are observed for the ferrimagnetic hcp planes for eutectic stable concentrations in the range $0.23<c<0.5$.

\subsection{Alloy layered $[Co_{1-c}Gd_{c}]_3[Co]_3[Co_{1-c}Gd_{c}]_3$ ferrimagnetic nanostructure between cobalt leads}
\label{333}
The layered $[3^{\prime}33^{\prime}]$ nanostructures under consideration are symmetric about the origin here taken as the hcp plane $n=0$. In this system the symmetry properties to be used are
\begin{eqnarray}
\left(
\begin{array}{cc}
	\sigma^{(1)}_{Co} \\
	\sigma^{(1)}_{Gd} \\
\end{array}\right)&\equiv&
\left(
\begin{array}{cc}
	\sigma^{(9)}_{Co}  \\
	\sigma^{(9)}_{Gd}  \\
\end{array}\right),\left(
\begin{array}{cc}
	\sigma^{(2)}_{Co} \\
	\sigma^{(2)}_{Gd} \\
\end{array}\right)\equiv
\left(
\begin{array}{cc}
	\sigma^{(8)}_{Co}  \\
	\sigma^{(8)}_{Gd}  \\
\end{array}\right),\nonumber\\
\left(
\begin{array}{cc}
	\sigma^{(3)}_{Co} \\
	\sigma^{(3)}_{Gd} \\
\end{array}\right)&\equiv&
\left(
\begin{array}{cc}
	\sigma^{(7)}_{Co}  \\
	\sigma^{(7)}_{Gd}  \\
\end{array}\right), \text{and}
\left(
\begin{array}{cc}
	\sigma^{(4)}_{Co} \\
\end{array}\right)\equiv
\left(
\begin{array}{cc}
	\sigma^{(6)}_{Co}  \\
\end{array}\right)
\end{eqnarray}

Similarly, as in the previous case $[2^{\prime}22^{\prime}]$, the equivalent molecular field energy results can be cast here in matrix form as

\begin{eqnarray}
\label{nonlineareq333}
& &\left(
\begin{array}{c}
	H_{Co}^{(1)}  \\
	H_{Gd}^{(1)}  \\
	H_{Co}^{(2)}  \\
	H_{Gd}^{(2)}  \\
	H_{Co}^{(3)}  \\
	H_{Gd}^{(3)}  \\
	H_{Co}^{(4)}  \\
	H_{Co}^{(5)}  \\
\end{array}\right)=
\left(
\begin{array}{c}
	A_1  \\
	A_2\\
	0  \\
	0\\
	0\\
	0\\
	0\\
	0\\
\end{array}\right)\nonumber\\&+&
\left(
\begin{array}{cccccccc}
	2 \beta J_{cc} & 6cJ_{cg} & \beta J_{cc}  & 3cJ_{cg} & 0 & 0 & 0 & 0  \\
	2 \beta J_{cg} & 6cJ_{gg} & \beta J_{cg}  & 3cJ_{gg} & 0 & 0 & 0 & 0 \\
	\beta J_{cc} & 3cJ_{cg} & 2\beta J_{cc}  & 6cJ_{cg} & \beta J_{cc} & 3c J_{cg} & 0 &0   \\
	\beta J_{cg} & 3cJ_{gg} & 2\beta J_{cg}  & 6cJ_{gg} & \beta J_{cg} & 3 c J_{gg} & 0 & 0  \\
	0 & 0 & \beta J_{cc}  & 3cJ_{cg} & 2\beta J_{cc}& 6c J_{cg} & 3 J_{cc} & 0  \\
	0 & 0 & \beta J_{cg}  & 3cJ_{gg} & 2\beta J_{cg}& 6c J_{gg} & 3 J_{cg} & 0  \\
	0 & 0 & 0 & 0 & \beta J_{cc} & 3c J_{cg} & 6 J_{cc} & 3 J_{cc} \\
	0 & 0 & 0 & 0 & 0 & 0 & 6 J_{cc} & 6 J_{cc} \\
\end{array}\right)\nonumber\\
&\times&
\left(
\begin{array}{c}
	\sigma^{(1)}_{Co} \\
	\sigma^{(1)}_{Gd} \\
	\sigma^{(2)}_{Co} \\
	\sigma^{(2)}_{Gd} \\
	\sigma^{(3)}_{Co} \\
	\sigma^{(3)}_{Gd}\\
	\sigma^{(4)}_{Co}\\
	\sigma^{(5)}_{Co}\\
\end{array}\right)
\end{eqnarray}
where $A_1(kT)=3 M_{Co}(kT) J_{Co-Co}$ and $A_2(kT)=3 M_{Co}(kT) J_{Co-Gd}$. This yields the new irreducible variables

\begin{eqnarray}
\left(
\begin{array}{c}
	\sigma^{(1)}_{Co} \\
	\sigma^{(1)}_{Gd} \\
	\sigma^{(2)}_{Co} \\
	\sigma^{(2)}_{Gd} \\
	\sigma^{(3)}_{Co} \\
	\sigma^{(3)}_{Gd}\\
	\sigma^{(4)}_{Co}\\
	\sigma^{(5)}_{Co}\\
\end{array}\right)=
\left(
\begin{array}{c}
	B_{Co}(S_{Co},T,H_{Co}^{(1)})  \\
	B_{Gd}(S_{Gd},T,H_{Gd}^{(1)})  \\
	B_{Co}(S_{Co},T,H_{Co}^{(2)})  \\
	B_{Gd}(S_{Gd},T,H_{Gd}^{(2)})  \\
	B_{Co}(S_{Co},T,H_{Co}^{(3)})  \\
	B_{Gd}(S_{Gd},T,H_{Gd}^{(3)})  \\
	B_{Co}(S_{Co},T,H_{Co}^{(4)})  \\
	B_{Co}(S_{Co},T,H_{Co}^{(5)})  \\
\end{array}\right)
\end{eqnarray}
\begin{figurehere}
	\centering
	\includegraphics[scale=0.67]{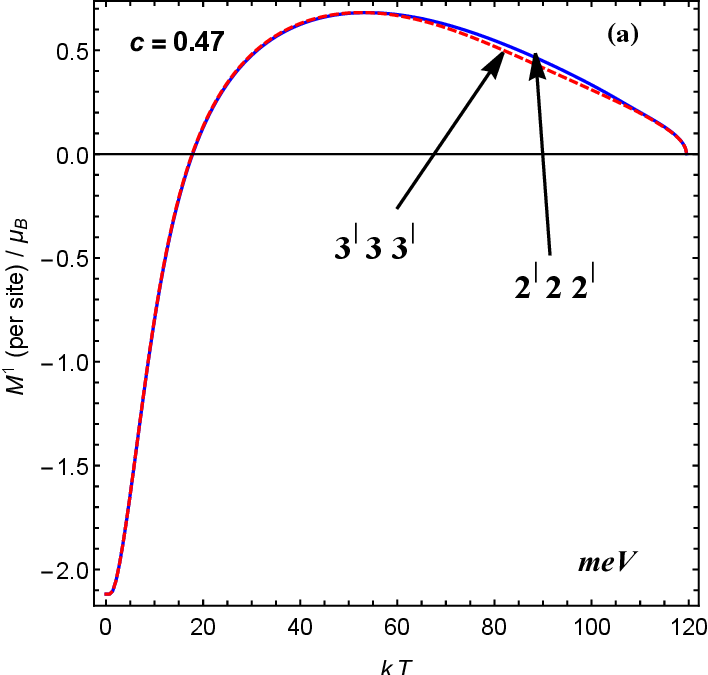}
	\includegraphics[scale=0.63]{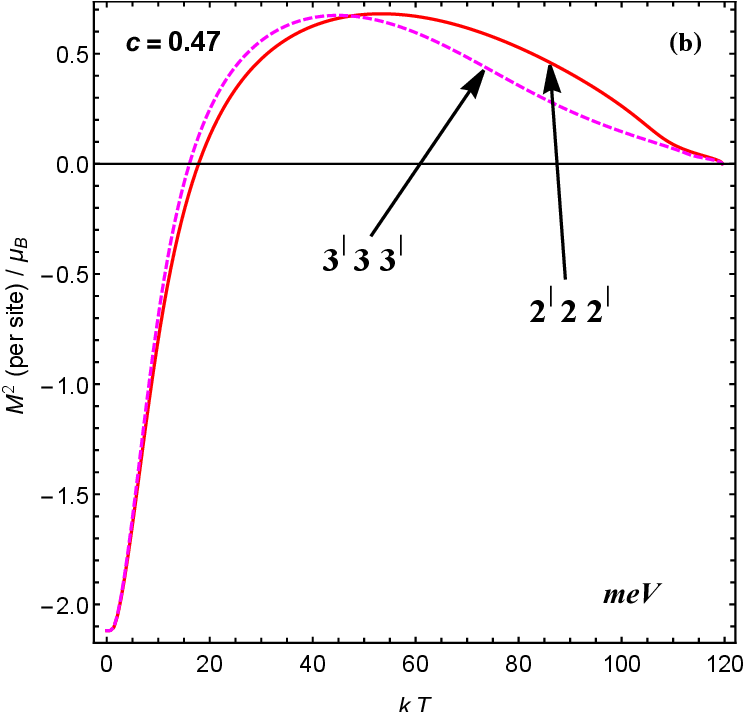}
	\caption{Calculated effective magnetic moments per site for the nominal: (a) 1st alloyed, ${n^{\prime}}$ = 1, and (b) 2nd alloyed, ${n^{\prime}} = 2$, hcp basal planes, for the alloy layered $[Co_{0.53}Gd_{0.47}]_2  [Co]_2 [Co_{0.53}Gd_{0.47}]_2$\ and  $[Co_{0.53}Gd_{0.47}]_3[Co]_3 [Co_{0.53}Gd_{0.47}]_3$\ nanostructures between cobalt leads.}\label{222333Mmomentpersite1}
\end{figurehere}
The above nonlinear equations can be solve numerically to obtain the spin variables and magnetic moments per site, on the individual hcp basal planes of the alloy layered magnetic $[3^{\prime}33^{\prime}]$ nanostructure. Fig.\ref{333Mmomentpersite123} presents the calculated results for the magnetic moments on the 1st, 2nd, and 3rd hcp atomic planes for this system. The magnetic moments per site for the 1st and 3rd hcp planes (discontinuous and continues curves, respectively) are quite similar throughout the temperature range of the ordered phase. Together, they differ significantly from the magnetic moments per site for the 2nd hcp plane, throughout the temperature range of the ordered phase.

As defined, the fundamental atomic spins are $S = 1$ for $Co$, and $S = 7/2$ for $Gd$, where we consider the spins to be up for $Co$ and down for $Gd$, in the ferrimagnetic alloy. Note that the $Gd$ fundamental spin is 3.5 times that of $Co$. However, the magnetic exchange $J_{gg}$ between $Gd$ and $Gd$ is weaker than $J_{cc}$ between $Co$ and $Co$. By studying equations 7, 8, 9, and 10, one can see that the molecular fields for $Co$ and $Gd$ sites vary with temperature and can change signs. The competition between these molecular fields along the entire temperature range determines the magnetizations per site, $M$; in the low temperature regime they increase then reach a maximum value before losing their magnetized phase. See figures 3 and 4. 

The computational model is general and can be extended to treat individual hcp atomic planes of alloy layered  $[Co_{1-c}Gd_{c}]_p[Co]_q[Co_{1-c}Gd_{c}]_r$ nanostructures, with $r, p, q\geq 1$. This procedure may be generalized to larger layered nanostructures between semi-infinite $Co$ leads. It has been observed that while increasing  $r,p,q$ the results for the sublattice magnetization properties in the core atomic planes tend to limiting solutions.

\begin{figurehere}
\centering
\includegraphics[scale=0.75]{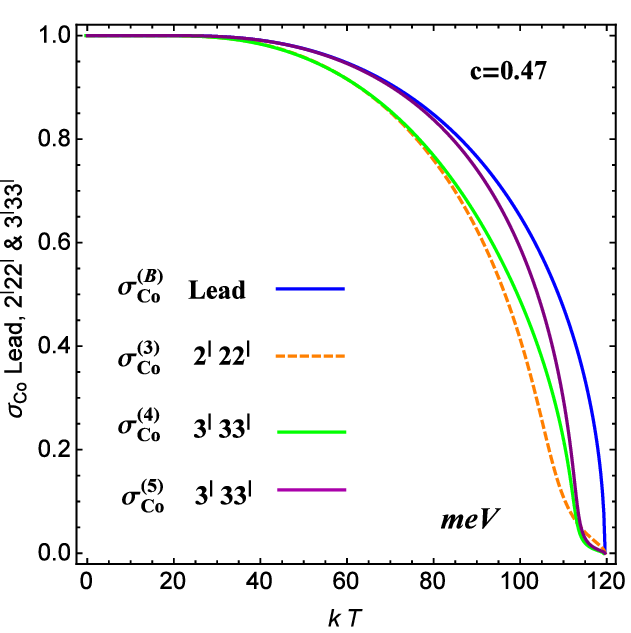}
\caption{Calculated spin variables $\sigma_{Co}^{(n^{\prime})}$ and their detailed variations as a function of temperature  on the nominal ${n^{\prime}}$ cobalt hcp basal planes inside the alloy layered ferrimagnetic nanostructures $[Co_{0.53}Gd_{0.47}]_2  [Co]_2 [Co_{0.53}Gd_{0.47}]_2$\ and  $[Co_{0.53}Gd_{0.47}]_3  [Co]_3 [Co_{0.53}Gd_{0.47}]_3$\, in comparison with the  temperature variation of the spin variable  $\sigma_{Co}^{(B)}$  on the cobalt leads; see details in the text.}\label{Spinvarlead222333}
\end{figurehere}


\begin{table*}
	\tbl{Spin variable values $\sigma_{Gd}=<S_{Gd}>$ for the $Gd$ sites on atomic planes 1, 2, .., $\ell$, of the $2^{\prime}22^{\prime}$ and $3^{\prime}33^{\prime}$  nanojunction systems between cobalt leads, are given for stable eutectic compositions $c\leq$0.5, at room temperature T=300K, using the theoretical EFT-MFT combined method. The spin variable $\sigma_{Co}=<S_{Co}>$ is $\approx1$ at room temperature for the $Co$ sites throughout the system.}
	{\begin{tabular}{@{}cccccccccccc@{}}
\toprule
\multicolumn{2}{c}{Concentrations}&\multicolumn{10}{c}{Spin variables $\sigma_{Gd}$}\\
\multicolumn{2}{c}{}&\multicolumn{4}{c}{$2^{\prime}22^{\prime}$}&\multicolumn{6}{c}{$3^{\prime}33^{\prime}$ }\\
\cline{3-12}
\multicolumn{2}{c}{$c$}&\multicolumn{2}{c}{$[Co_{1-c}Gd_c]_{2,L}$ }&\multicolumn{2}{c}{$[Co_{1-c}Gd_c]_{2,R}$ }&\multicolumn{3}{c}{$[Co_{1-c}Gd_c]_{3,L}$}&\multicolumn{3}{c}{$[Co_{1-c}Gd_c]_{3,R}$}\\
\colrule
&0.1& -1.10&-1.10 &-1.10 &-1.10 &-1.10 & -1.08&-1.10 &-1.10 &-1.08 &-1.10\\
&0.2& -1.03&-1.03 & -1.03&-1.03 &-1.03 &-1.00 &-1.03 &-1.03 &-1.00 &-1.03\\
&0.3& -0.97& -0.97&-0.97 &-0.97 &-0.97 & -0.91&-0.97 &-0.97 &-0.91 &-0.97\\
&0.4&-0.91 &-0.91 & -0.91& -0.91&-0.90 &-0.81 &-0.90 &-0.90 &-0.81 &-0.90\\
&0.5& -0.83&-0.83 & -0.83&-0.83 &-0.82 &-0.71 &-0.82 &-0.82 &-0.71 &-0.82\\
\botrule
		\end{tabular}\label{SpinvariableofGd}
	}
	
\end{table*}

\subsection{Alloy layered\, $[Co_{0.53}Gd_{0.47}]_{\ell^{\prime}} [Co]_\ell [Co_{0.53}Gd_{0.47}]_{\ell^{\prime}}$\, ferrimagnetic nanostructure between cobalt leads}

This alloy layered magnetic nanostructure between cobalt leads, at the characteristic eutectic concentration $c = 0.47$, is particularly interesting since $Co/Gd$ magnetic multilayers at the same composition have been reported to be very stable, \cite{Andres08}. We have applied hence the EFT - MFT model approach to deduce the spin variables $\sigma_{Co}^{(n^{\prime})}$ and $\sigma_{Gd}^{(n^{\prime})}$, and the effective magnetic moments per site, for the individual hcp basal planes of layered ferrimagnetic nanostructures between cobalt leads, as a function of temperature, eutectic concentration, and thicknesses $\ell$ = 2 and 3. The integer $n^{\prime}$ numbers the hcp planes from 1 to 6 for $\ell=2$, and from 1 to 9 for $\ell=3$.

The  effective magnetic moments per site calculated in the units of Bohr magnetons for the alloyed nominal 1st, ${n^{\prime}}$ = 1, and 2nd, ${n^{\prime}}$ = 2, hcp basal planes for the alloy layered ferrimagnetic nanostructures $[Co_{0.53}Gd_{0.47}]_2  [Co]_2 [Co_{0.53}Gd_{0.47}]_2$\ and  $[Co_{0.53}Gd_{0.47}]_3  [Co]_3 [Co_{0.53}Gd_{0.47}]_3$\, are presented in Fig.\ref{222333Mmomentpersite1}.
It is observed that the computed effective magnetic moments per site as a function of temperature on the  nominal ${n^{\prime}}$ = 1 hcp basal planes do not vary significantly with increased thickness of nanostructure, see  Fig.\ref{222333Mmomentpersite1}(a). This can be understood clearly since the corresponding matrix elements in Eq.(\ref{nonlineareq333}) do not change significantly with increasing thickness. In contrast, it is observed that the effective magnetic moments per site on the nominal ${n^{\prime}}$ = 2 hcp basal planes do vary significantly with increasing thickness, see  Fig.\ref{222333Mmomentpersite1}(b). This is expected physically owing to the changes of the corresponding effective molecular fields for $Co$ and $Gd$ sites. The observed variations start at $\approx$15 meV and persist for higher temperatures, including room temperature $\approx$26 meV.

It is also interesting to compute spin variables $\sigma_{Co}^{(n^{\prime})}$, and their detailed variations as a function of temperature  on the nominal ${n^{\prime}}$ cobalt hcp basal planes inside the alloy layered ferrimagnetic nanostructures, in comparison with the spin variable $\sigma_{Co}^{(B)}$ on the cobalt leads as a function of temperature. This is done for the $[Co_{0.53}Gd_{0.47}]_{\ell^{\prime}} [Co]_\ell [Co_{0.53}Gd_{0.47}]_{\ell^{\prime}}$ layered nanostructure between cobalt leads, for thicknesses $\ell$ = 2 and 3. The calculated results are presented in Fig.\ref{Spinvarlead222333}. As is physically expected, the $\sigma_{Co}^{(B)}$ is $\geq \sigma_{Co}^{(n^{\prime})}$  for all  pure $Co$ hcp planes ${n^{\prime}}$ inside the layered thicknesses $\ell$, at all temperatures of the ordered ferrimagnetic phase. Also as expected, $\sigma_{Co}^{(5)}$ is  $\geq \sigma_{Co}^{(4)}$ for the alloy layered $[Co_{0.53}Gd_{0.47}]_3  [Co]_3 [Co_{0.53}Gd_{0.47}]_3$ ferrimagnetic nanostructure between cobalt leads, and both are greater or equal to the $\sigma_{Co}^{(3)}$  of the $[Co_{0.53}Gd_{0.47}]_2  [Co]_2 [Co_{0.53}Gd_{0.47}]_2$ layered nanostructure. Note that the $n^{\prime}=3$ pure cobalt plane for the $[2^{\prime}22^{\prime}]$ layered  nanostructure is equivalent nominally to the $n^{\prime}=4$ pure cobalt plane for the $[3^{\prime}33^{\prime}]$ layered nanostructure. The results confirm a physical trend which is expected, and which would lead to limiting values with increasing thickness of the layered ferrimagnetic structure between cobalt leads.

We emphasize that the basic physical variables, such as the exchange and sublattice magnetizations for $Co$  and $Gd$ sites for the embedded layered nanostructures between cobalt leads, are necessary elements for the computations of the spin-dynamics of magnetic nanojunctions in the field of magnonics, as for the ballistic magnon transport \cite{Ashokan15}. In Table \ref{SpinvariableofGd} we present an example for the calculated spin variables $<S_{Co}>$ $\approx1$ throughout the system, and $<S_{Gd}>$ on the identified atomic planes 1, 2, .., $\ell$, of the layered $...Co][2^{\prime}22^{\prime}][Co...$ and $...Co][3^{\prime}33^{\prime}][Co...$  nanostructures between cobalt leads, at room temperature T=300K and stable eutectic compositions $c\leq$ 0.5, using the EFT-MFT combined method.

\section{Summary and conclusions}
\label{Conclusions}

In this work, we model the salient sub-lattice magnetic properties of the alloy layered ferrimagnetic $[Co_{1-c}Gd_c]_{\ell^{\prime}}[Co]_\ell[Co_{1-c}Gd_c]_{\ell^{\prime}}$ nanostructures between magnetically ordered cobalt leads. In particular, sublattice magnetizations of the $Co$  and $Gd$ sites on the individual hcp (0001) basal planes of the alloy layered lamellar  nanostructures by using EFT-MFT combined method. The  effective magnetic moments per site and sublattice magnetizations are plotted as a function of  temperature and thicknesses of the lamellar nanostructure. The computational model is general and can represents other composite magnetic elements and lamellar nanostructures.
The calculated magnetic exchange, and spin variables for gadolinium $<S_{Gd}>$ and cobalt $<S_{Co}>$, site on the identified hcp (0001) basal planes of the alloy layered ferrimagnetic $[Co_{1-c}Gd_c]_{\ell^{\prime}}[Co]_\ell[Co_{1-c}Gd_c]_{\ell^{\prime}}$  nanostructures between cobalt leads, are very important quantities for the self-consistent analysis of  quantum spin dynamics  system and the coherent magnon ballistic transport  across such nanostructures. The calculated results are also important fo the  applications in the fields of magnonics. The Ising EFT method serves to determine  the magnetic exchange constants for $Co$ and $Gd$ sites of pure crystals, characterized by their fundamental quantum spins, by comparing  EFT  with the  experimental data. By seeding the MFT results on the alloy lamellar ferrimagnetic  nanostructure by the EFT computations of the cobalt leads from the interface inwards, the sublattice magnetizations for the  $Co$  and $Gd$  sites in the nanostructure are computed.



\end{multicols}

\begin{thebibliography}{36}
\label{thebibliographie}
\bibitem{Serga07} A.A. Serga, A.V. Chumak, A. Andre, G.A. Melkov, A.N. Slavin, S.O. Demokritov, and B. Hillebrands, Phys. Rev. Lett. {\bf99}, (2007) 227202
\bibitem{Schneider08} T. Schneider, A.A. Serga, B. Leven, B. Hillebrands, R.L. Stamps, and M.P. Kostylev, Appl. Phys. Lett. {\bf92}, (2008) 022505
\bibitem{Kruglyak10} V. V. Kruglyak, S. O. Demokritov and D. Grundler, J. Phys. D: Appl. Phys. {\bf43}, (2010) 264001
\bibitem{Lee08} K. Lee and S. Kim, J. Appl. Phys. {\bf104}, (2008) 053909
\bibitem{Wolf04} S. A. Wolf, D. D. Awschalom, R. A. Buhrman, J. M. Daughton, S. von Moln\'{a}r, M. L. Roukes, A. Y. Chtchelkanova and D. M. Treger , Science {\bf294}, (2004) 1488
\bibitem{Zutic04} I. Zutic, Jaroslav Fabian and S. Das Sarma, Rev. Mod. Phys. {\bf76},(2004) 323
\bibitem{Bogani08} L. Bogani and W. Wernsdofer, Nature Materials, {\bf7}, (2008) 179
\bibitem{Camley93} R.E. Camley and R. L. Stamps, J. Phys. Condens. Matter, {\bf5}, (1993) 3727
\bibitem{Andres00} J.P. Andr\'{e}s, J. L. Sacedo\'{n}, J. Colinoa) and J. M. Riveiro J. Appl. Phys. {\bf87}, (2000) 2483
\bibitem{Anilturk03} O.S. Anilturk and A.R. Koymen, Phys. Rev. B {\bf68}, (2003) 024430
\bibitem{Gonzalez04} J.A. Gonzalez, J. Colino, J.P. Andres, M.A. Lopez de la Torre, J.M. Riveiro, Physica B {\bf345}, (2004) 181
\bibitem{Andres08} J.P. Andr\'{e}s, J. A. Gonzalez,T. P. A. Hase, B. K. Tanner, and J. M. Riveiro, Phys. Rev. B {\bf77}, (2008) 144407
\bibitem{Demirtas05} S. Demirtas, R. E. Camley, A. R. Koymena, Appl. Phy. Lett. {\bf87}, (2005) 202111
\bibitem{Javier22} Javier Hermosa-Mu\"{n}oz at. al 
Communications Physics, {\bf 5}, (2022) 26.
 \bibitem{Chaudhari73}  P. Chaudhari, J. J. Cuomo, and R. J. Gambino, Appl. Phys. Lett. {\bf22},  (1973) 337
\bibitem{Hansen89} P. Hansen, C. Clausen, G. Much, M. Rosenkranz, and K. Witter, J. App. Phys. {\bf66},  (1989) 756
\bibitem{Khater92} A. Khater, G. Le Gal, and T. Kaneyoshi, Phys. Letters A {\bf171},  (1992) 237
\bibitem{Fresneau94} M. Fresneau, G. Le Gal, and A. Khater, J. Mag. Mag. Mat. {\bf130},  (1994) 63
\bibitem{Khater021} A. Khater, M. Abou Ghantous, and M. Fresneau, J. Mag. and Mag. Mat. {\bf247},  (2002) 305
\bibitem{Camley88} R.E. Camley and D.R. Tilley,  Phys. Rev. B {\bf37},  (1988) 3413
\bibitem{Mansuripur86} M. Mansuripur and M.F. Ruane,  IEEE Trans. Magn. {\bf22},  (1986) 33
\bibitem{Vaz08} C.A.F. Vaz, J.A.C. Bland, and G. Lauhoff, Rep. Prog. Phys. {\bf71},  (2008) 056501
\bibitem{Ashokan14} V. Ashokan, M. Abou Ghantous, D. Ghader, and A. Khater, J. Mag. Mag. Mat.{\bf363},  (2014) 66
\bibitem{Ghantous13} M. Abou Ghantous, A. Khater, V. Ashokan, D. Ghader, J. Appl. Phys {\bf113},  (2013) 094303
\bibitem{Ashokan15}  V. Ashokan, A. Khater, M. Abou Ghantous, D. Ghader, J. Mag. Mag. Mat. {\bf384},  (2015) 18
\bibitem{Ashokan16} V. Ashokan, M. Abou Ghantous, D. Ghader, A. Khater, Thin Solid Films {\bf616} (2016) 6
\bibitem{Khater18} A. Khater, L. Saimb, R. Tigrineb, D. Ghader, Surface Science {\bf 672} 673 (2018) 47
\bibitem{Chelli20} Farid Chelli, Boualem Bourahla and Antoine Khater, Int. J. of Mod. Phys. B, {\bf34}, (2020) 2050080
\bibitem{Kuzmin05} M.D. Kuz'min, Phys. Rev. Lett. {\bf94},  (2005) 107204
\bibitem{Khater15} Elie A. Moujaes, A. Khater, M. Abou Ghantous, J. Mag. and Mag. Mat. 391 (2015) 49
\bibitem{Tucker94}J. W. Tucker,  J. Phys. A: Math. Gen. {\bf27},  (1994) 659
\bibitem{Khater11} A. Khater and M. Abou Ghantous,  J. Mag. Mag. Mat. {\bf323},  (2011) 2717
\bibitem{MAbouGhantous11} M. Abou Ghantous and  A. Khater,  J. Mag. Mag. Mat. {\bf323},  (2011) 2504
\bibitem{Honmura79} R. Honmura and T. Kaneyoshi,  J. Phys. C: Solid St. Phys. {\bf12},  (1979) 3979
\bibitem{Myers51} H.P. Myers, and W. Sucksmith, Proc. R. Soc. London A {\bf207},  (1951) 427
\bibitem{Pauthenet82} R. Pauthenet,  J. Appl. Phys. {\bf53}, (1982) 8187
\bibitem{Nigh63} H.E. Nigh, S. Legvold, and F.H. Spedding,  Phys. Rev. {\bf132},  (1963) 1092.



\end{thebibliography}
\end{document}